\documentclass[prc,twocolumn,nofootinbib,superscriptaddress,showpacs]{revtex4}
\usepackage{graphicx,amsmath,amssymb,bm,multirow}

\newcommand{\be}[1]{\begin{equation}\label{#1}}
\newcommand{\ee}{\end{equation}}
\newcommand{\vlowk}{V_{{\rm low}\,k}}
\newcommand{\lm}{\Lambda}
\newcommand{\jrel}{J_{\rm rel}}
\newcommand{\la}{\langle}
\newcommand{\ra}{\rangle}
\newcommand{\fm}{\, \text{fm}}
\newcommand{\fmi}{\, \text{fm}^{-1}}
\newcommand{\mev}{\, \text{MeV}}

\newcommand{\elemA}[2]{\ensuremath{{}^{#1}}\textrm{#2}}

\newcommand{\partialwave}[3]{\ensuremath{{}^{#1}}\textrm{#2}_{#3}}

\begin{document}

\title{Partial-wave contributions to pairing in nuclei} 

\author{Simone Baroni}
\email[E-mail:~]{baroni@phys.washington.edu}
\affiliation{Institute for Nuclear Theory,
University of Washington, Box 351550, Seattle, WA 98195, USA}
\affiliation{TRIUMF, 4004 Wesbrook Mall, Vancouver BC, V6T 2A3, Canada}
\author{Augusto O.\ Macchiavelli}
\email[E-mail:~]{aomacchiavelli@lbl.gov}
\affiliation{Nuclear Science Division, Lawrence Berkeley National Laboratory,
Berkeley, CA 94720, USA}
\author{Achim Schwenk}
\email[E-mail:~]{schwenk@triumf.ca}
\affiliation{TRIUMF, 4004 Wesbrook Mall, Vancouver BC, V6T 2A3, Canada}
\affiliation{ExtreMe Matter Institute EMMI, GSI Helmholtzzentrum f\"ur
Schwerionenforschung GmbH, 64291 Darmstadt, Germany}
\affiliation{Institut f\"ur Kernphysik, Technische Universit\"at
Darmstadt, 64289 Darmstadt, Germany}

\begin{abstract}
We present a detailed study of partial-wave contributions of nuclear
forces to pairing in nuclei. For $T=1, J=0$ pairing, partial waves
beyond the standard $\partialwave{1}{S}{0}$ channel play an
interesting role for the pair formation in nuclei. The additional
contributions are dominated by the repulsive $\partialwave{3}{P}{1}$
partial wave. Their effects, and generally spin-triplet nuclear
forces between paired nucleons, are influenced by the interplay of
spin-orbit partners. We explore the impact of including partial
waves beyond the $\partialwave{1}{S}{0}$ channel on neutron-neutron
pairing gaps in semi-magic isotopic chains. In addition, we show
that nuclear forces favor $T=1, J=0$ over $T=0, J=1$ pairing, except
in low-$j$ orbitals. This is in contrast to the free-space
motivation that suggests the formation of deuteron-like $T=0$ pairs
in $N=Z$ nuclei. The suppression of $T=0$ pairing is because the
$\partialwave{3}{S}{1}$ strength is distributed on
spin-orbit partners and because of the effects of the repulsive
$\partialwave{1}{P}{1}$ channel and of D waves.
\end{abstract}


\maketitle

\section{Introduction}
\label{intro}

Phenomenological energy density functionals are impressively
successful in the description of medium-mass and heavy
nuclei~\cite{Bender}, but lack a microscopic connection to nuclear
forces and seem to have reached the limits of improvement in the
present functional form~\cite{Bertsch,Kortelainen}. This has lead to
exciting efforts, largely driven by effective field theory (EFT)
ideas, to develop a universal energy density functional based on
microscopic interactions~\cite{DFTreview,Wolfram,DME,UNEDF,Biruk}. These
developments rely on the Hartree-Fock approximation as a good starting
point for nuclei or on a perturbative expansion about nuclear
matter. Both have become possible by evolving nuclear forces to
low-momentum interactions using the renormalization group
(RG)~\cite{Vlowk,smooth,SRG,nucmatt,chiralnm,neutmatt}.

In this paper, we follow Refs.~\cite{Duguet1,Lesinski} and use
existing energy density functionals in the particle-hole channel, to
build a reasonable self-consistent single-particle basis, combined with
low-momentum interactions $\vlowk$~\cite{Vlowk,smooth} in the pairing
channel. The calculational details are given in Sect.~\ref{calc}. This
was shown to provide a good starting description of neutron-neutron
and proton-proton pairing properties in semi-magic
nuclei~\cite{Duguet1,Lesinski,Duguet2,Hebeler}. In these and
other~\cite{Barranco,Pastore} calculations based on nuclear forces,
the pairing interaction was restricted to the $\partialwave{1}{S}{0}$
partial wave. This is motivated by pairing in infinite matter (see,
for example Ref.~\cite{Kai}), but in nuclei the paired nucleons are
not in back-to-back-momentum configurations. Therefore, other partial
waves can contribute to the pairing interaction for two particles with
isospin $T$ and total angular momentum $J$.

In Sect.~\ref{pw}, we discuss partial-wave contributions to pairing
interactions in nuclei. This is followed in Sect.~\ref{me} by a
detailed analysis of $T=1, J=0$ pairing matrix elements in the $sd$
and $pf$ shells, where we focus on the effects of partial waves beyond
the standard $\partialwave{1}{S}{0}$ channel. Spin-triplet
partial-wave contributions to pairing in time-reversed states are
found to depend on the spin-orbit configurations involved. In
Sect.~\ref{T1}, we explore the impact on $T=1$ pairing gaps in
semi-magic isotopic chains. Our analysis of pairing matrix elements
provides a simple explanation of the results observed in nuclei.

We extend the study of partial-wave contributions to $T=0$ pairing in
Sect.~\ref{T0}, and compare the pairing strengths to the $T=1, J=0$
channel at the level of low-momentum interactions. Since the deuteron
is bound, the free-space motivation suggests the formation of $T=0,
J=1$ pairs in all $N=Z$ nuclei~\cite{T0ref}. In contrast, there is no
direct observation of $T=0$ pairing to date. While we do not perform a
calculation for nuclei, our analysis of $T=0, J=1$ pairing matrix
elements based on nuclear forces can provide a microscopic explanation
for the suppression of $T=0$ pairing. For the $\partialwave{3}{S}{1}$
channel, the nuclear force strength is distributed on the spin-orbit
partners, leading to a geometrical suppression in higher-$j$
orbitals. Moreover, the additional partial-wave contributions are
dominated by the repulsive $\partialwave{1}{P}{1}$ channel with some
smaller contributions due to D waves. As a result, in the absence of
many-body effects on pairing, we find that low-momentum interactions
favor $T=1, J=0$ over $T=0, J=1$ pairing, except in light $N=Z$ nuclei
where pairing is dominated by low-$j$ orbitals. The standard
motivation for $T=0$ pairing in nuclei is therefore at best
incomplete. We conclude and give an outlook in Sect.~\ref{concl}.

\section{Calculational details}
\label{calc}

The minimization of the energy density functional in the presence of
pairing leads to solving the self-consistent Hartree-Fock-Bogoliubov (HFB)
equations~\cite{Bender,RingSchuck} that determine the quasiparticle
(q) basis,
\be{HFB}
\left(\begin{array}{cc}
h-\mu & \Delta \\
-\Delta^* & -(h-\mu) \end{array}\right)
\left(\begin{array}{c}
U^q \\ V^q \end{array}\right)
= E_q \left(\begin{array}{c} U^q \\ V^q \end{array}\right) \,.
\ee
Here $h$ denotes the single-particle Hamiltonian, $\mu$ the Fermi
level, $E_q$ is the quasiparticle energy and $U^q, V^q$ the
corresponding coefficients of the Bogoliubov transformation from
single-particle to quasiparticle states. We use the Skyrme functional
SLy4~\cite{SLy4} and first solve the Hartree-Fock (HF) equations,
$h \phi_a = \varepsilon_a^{\rm HF} \phi_a$, on a spherical mesh with $0.1 \fm$
spacing and $16.0 \fm$ box radius. Our results are stable
with respect to increasing the radius and decreasing the
mesh spacing. This defines the single-particle
basis $| a \ra$, using the shorthand label $a \equiv n_a l_a j_a$
with radial quantum number $n_a$, orbital angular momentum $l_a$
and total angular momentum $j_a= l_a \pm 1/2$.

Using the HF single-particle Hamiltonian, we then solve the HFB
equations, where the state-dependent gap matrix $\Delta$ for $T=1,
J=0$ pairing is given by
\begin{align}
\Delta_{ab} &= -\sum_{cd} \, \biggl( \sum_q U^q_c V^q_d \biggr)
\sqrt{\frac{2j_c+1}{2j_a+1}} \, \frac{\sqrt{(1+\delta_{ab})
(1+\delta_{cd})}}{2} \nonumber \\[1mm]
& \times \la a \, b \, | \, (1 - P_{12}) \, \vlowk \, | \, c \, d
\, \ra_{J=0, T=1, T_z=-1} \,.
\label{gapeq}
\end{align}

This defines the gap equation and we focus on neutron-neutron ($T_z =
-1$) pairing properties. The matrix elements of the pairing
interaction in the second line of Eq.~(\ref{gapeq}) are
antisymmetrized using the exchange operator $P_{12}$ and normalized,
and $\delta_{ab}$ is shorthand for $\delta_{n_a n_b} \, \delta_{l_a
l_b} \, \delta_{j_a j_b}$. In general, after the HFB equations are
solved self-consistently, one has to insert the resulting densities
back into the single-particle Hamiltonian, which leads to a new set of
single-particle states, in turn to a new HFB solution, and these
iterations have to be repeated to obtain the fully self-consistent HFB
solution. However, as shown in Sect.~\ref{T1}, the feedback on the
single-particle Hamiltonian can be neglected for pairing properties to
a good approximation.

For the neutron-neutron pairing interaction, we start from the chiral
N$^3$LO two-nucleon (NN) potential ($\lm=500 \mev$) of
Ref.~\cite{N3LO} and use the RG to evolve this NN potential to
low-momentum interactions $\vlowk$ with a smooth $n_{\rm exp}=4$
regulator with $\lm = 1.8-2.8 \fmi$~\cite{smooth}. This evolution
renders the many-body calculation more
controlled~\cite{nucmatt,chiralnm,neutmatt} and provides a good
starting point for connecting energy density functionals to nuclear
forces~\cite{DFTreview,Hebeler}. Based on the universality of
$\vlowk$~\cite{smooth,chiralnm}, we do not expect large differences
starting from different N$^3$LO potentials. As discussed in the
following section, we calculate the $jj$-coupled pairing matrix
elements entering the gap equation, Eq.~(\ref{gapeq}), by expanding
the HF single-particle states on the harmonic oscillator (HO)
basis. For pairing properties around the Fermi level, in particular
for the lowest-quasiparticle-energy canonical state and average gaps
presented in Sect.~\ref{T1}, we have found that the HFB
single-particle space can be restricted to states below $60 \mev$ (for
hard potentials with large cutoffs, states up to $\sim 1 \, {\rm
GeV}$ are needed for convergence~\cite{Barranco,Pastore}).  Finally,
an important direction for future work is to use energy density
functionals in the particle-hole channel that are based on the same
nuclear forces and to include many-body contributions to the pairing
interaction. First results in this direction have been presented in
Ref.~\cite{Hergert}.

In HFB theory, the interacting particles pair in orbitals related by
time-reversal symmetry~\cite{BM}, where the radial quantum numbers can
differ due to the Bogoliubov transformation~\cite{RingSchuck}.
Semiclassically, this is realized by two particles moving in the same
orbit with opposite time order, so that $(l_a,j_a)=(l_b,j_b)$ and
$(l_c,j_c)=(l_d,j_d)$ and hence the pairs have positive parity. For
completeness, the gap equation is in the so-called phase convention
II~\cite{BM}, for which the operation of the time-reversal operator
$\mathcal{T}$ on a single-particle state is given by $\mathcal{T} \, |
l_a j_a m_a \ra = (-1)^{j_a - m_a} \, | l_a j_a -m_a \ra$, with
magnetic quantum number $m_a$.

\section{Partial-wave contributions to pairing interactions in nuclei}
\label{pw}

The $jj$-coupled pairing matrix elements in the HF basis are obtained
from the interaction matrix elements of $\vlowk$ in two-particle 
spherical HO states, which are calculated from the different partial-wave
contributions using the standard recoupling formula,
\begin{align}
&\la a \, b \, | \, (1 - P_{12}) \, \vlowk \, | \, c \, d
\, \ra_{J, T, T_z} \bigl|_\text{HO basis} \nonumber \\[2mm]
&= \sum \limits_{\begin{array}{c} n,n',l,l',N,L \\
\lambda,\lambda',S,\jrel \end{array}}
\frac{\widehat{j_a} \, \widehat{j_b} \, \widehat{j_c} \,
\widehat{j_d} \, \widehat{\lambda}^{\, 2} \, \widehat{\lambda'}^2 \,
\widehat{S}^{\, 2} \, \widehat{J}^{\, 2}_{\rm rel}}
{\sqrt{(1+\delta_{ab})(1+\delta_{cd})}}
\left\{ \begin{array}{ccc}
L & l & \lambda \\
S & J & \jrel 
\end{array} \right\} \nonumber \\
&\times \left\{ \begin{array}{ccc}
L & l' & \lambda' \\
S & J & \jrel 
\end{array} \right\} 
\left\{ \begin{array}{ccc}
l_a & 1/2 & j_a \\
l_b & 1/2 & j_b \\
\lambda & S & J 
\end{array} \right\}
\left\{ \begin{array}{ccc}
l_c & 1/2 & j_c \\
l_d & 1/2 & j_d \\
\lambda' & S & J 
\end{array} \right\} \nonumber \\[2mm]
& \times \la n l N L | n_a l_a n_b l_b \lambda \ra \,
\la n' l' N L | n_c l_c n_d l_d \lambda' \ra \, (-1)^{\lambda +
\lambda'} \nonumber \\[2mm]
& \times \bigl( 1-(-1)^{l+S+T} \bigr) \,
\la n l | \vlowk^{T, T_z} | n' l' S \jrel \ra \,.
\label{meform} 
\end{align}
Here we use $\widehat{x}=\sqrt{2x+1}$ and standard notation for $6j$
and $9j$ symbols~\cite{angular}. The bracket $\la n l N L | n_a l_a
n_b l_b \lambda \ra$ denotes Talmi-Moshinsky
brackets~\cite{Kamuntavicius}, with relative and center-of-mass radial
and orbital quantum numbers $n,l,N,L$, and total orbital angular
momentum of the pair $\lambda$. The two-body spin is given by $S$ and
${\bm J}_{\rm rel} = {\bm l} + {\bm S}$ denotes the relative total
angular momentum. We use the standard notation $^{2S+1}l_{J_{\rm rel}}$
to denote the different partial waves, which we list for completeness:
\begin{equation}
\begin{array}{lcccccc}
\text{for $T=1$:} & \, \partialwave{1}{S}{0} \, &
\, \partialwave{3}{P}{0,1,2} \, &
\, \partialwave{1}{D}{2} \, &
\, \partialwave{3}{F}{2,3,4} \, &
\, \partialwave{1}{G}{4} \, & \dots \nonumber \\
&&&&&& \nonumber \\
\text{for $T=0$:} & \, \partialwave{3}{S}{1} \, &
\, \partialwave{1}{P}{1} \, &
\, \partialwave{3}{D}{1,2,3} \, &
\, \partialwave{1}{F}{3} \, &
\, \partialwave{3}{G}{3,4,5} \, & \dots \nonumber
\end{array}
\end{equation}

Additional restrictions come into play in the pairing problem when the
total angular momentum of the paired nucleons is $J=0$. Because ${\bm
L} + {\bm J}_{\rm rel} = {\bm J}$, one therefore has $L = J_{\rm
rel}$ for $J=0$. Moreover, because the parity of the pair is
positive, it follows that the relative and
center-of-mass orbital angular momentum $l$ and $L$ must have the same
parity. As a result, the relative orbital and total angular momentum
$l$ and $J_{\rm rel}$ also have the same parity, and therefore only
uncoupled channels $^{2S+1}l_{J_{\rm rel}=l}$ contribute
to $T=1, J=0$ pairing matrix elements. On the other hand, $T=0, J=1$
matrix elements are not affected by these additional constraints and
coupled channels also contribute to the pair formation in nuclei.

In summary, this shows that the $\partialwave{1}{S}{0}$ and
$\partialwave{3}{S}{1}-\partialwave{3}{D}{1}$ partial waves are not
the only channels contributing to $T=1$ and $T=0$ pairing in
nuclei. In addition, the following partial waves are part of the
pairing interaction:
\begin{equation}
\begin{array}{lcccccc}
\text{$T=1, J=0$:} & \, \partialwave{1}{S}{0} \, &
\, \partialwave{3}{P}{1} \, &
\, \partialwave{1}{D}{2} \, &
\, \partialwave{3}{F}{3} \, &
\, \partialwave{1}{G}{4} \, & \dots \nonumber \\
&&&&&& \nonumber \\
\text{$T=0, J=1$:} & \, \partialwave{3}{S}{1} \, &
\, \partialwave{1}{P}{1} \, &
\, \partialwave{3}{D}{1,2,3} \, &
\, \partialwave{1}{F}{3} \, &
\, \partialwave{3}{G}{3,4,5} \, & \dots \nonumber
\end{array}
\end{equation}
Interestingly, both $\partialwave{3}{P}{1}$ and
$\partialwave{1}{P}{1}$ partial waves are repulsive at the relevant
energies~\cite{PWA}. We therefore expect a reduction of the pairing
gap compared to studies based on the standard S-wave interactions.

\section{$T=1, J=0$ pairing matrix elements}
\label{me}

\begin{figure}[t]
\begin{center}
\includegraphics[clip=,scale=0.35]{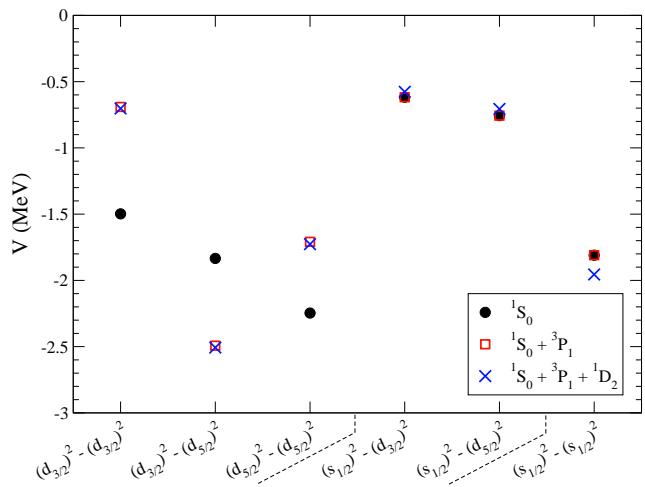}
\end{center}
\caption{$T=1, J=0$ pairing matrix elements in the $sd$ shell. 
Results are shown for the RG-evolved N$^3$LO potential of 
Ref.~\cite{N3LO} with $\lm = 2.0 \fmi$, represented in two-particle
HO states with $\hbar \omega = 10 \mev$. The label
$(l_{a\,j_a})^2-(l_{b\,j_b})^2$ denotes the bra$-$ket quantum numbers
and we have grouped matrix elements according to $(l_a,l_b)$. In the 
$sd$ shell, the two-particle quantum numbers combined with $J=0$
restrict the summation over partial waves to $l \leqslant 2$.
\label{sdT1}}
\end{figure}

\begin{figure}[t]
\begin{center}
\includegraphics[clip=,scale=0.35]{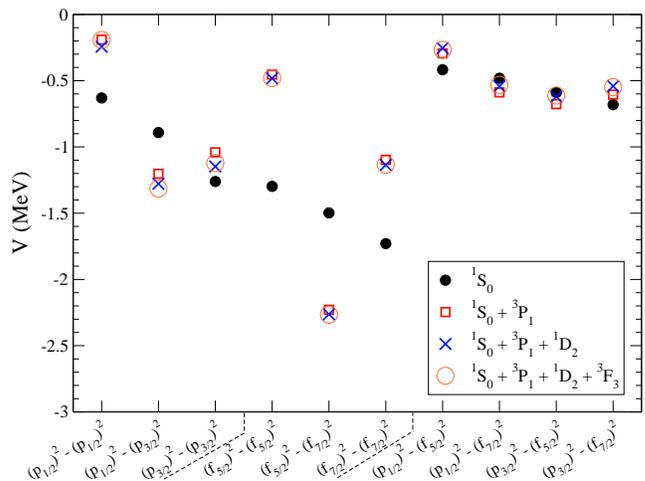}
\end{center}
\caption{Same as Fig.~\ref{sdT1} but for the $pf$ shell, where partial
waves with $l \leqslant 3$ can contribute to $T=1, J=0$ pairing matrix
elements.\label{pfT1}}
\end{figure}

The $sd$ and $pf$ shells are very useful for understanding the results
observed in nuclei in Sect.~\ref{T1}. These two major shells are small
enough to allow a simple study and at the same time they are complex
enough to draw general conclusions. In Figs.~\ref{sdT1}
and~\ref{pfT1}, we show the $jj$-coupled $T=1, J=0$ pairing matrix
elements in the $sd$ and $pf$ shells based on the smooth $\vlowk$
evolved from the N$^3$LO potential of Ref.~\cite{N3LO} to $\lm = 2.0
\fmi$. Here, we give neutron-proton ($T_z=0$) matrix elements in order
to compare to the $T=0, J=1$ pairing strengths in Sect.~\ref{T0}. The
contributions from isospin-dependent nuclear forces that distinguish
between the $T=1, T_z=0$ and $T=1, T_z=-1$ channel (used in the
calculations for nuclei) are small and do not change this analysis.
The matrix elements are in the HO basis with $\hbar \omega = 10 \mev$,
but the general effects of partial waves are qualitatively similar for
other values of $\hbar \omega$ or in the HF basis.

As expected based on NN phase shifts, the pairing matrix elements are
attractive for the $\partialwave{1}{S}{0}$ channel. In addition, we
find that the attractive $\partialwave{1}{S}{0}$ contribution
increases approximately linearly with increasing $j_a$ or $j_b$ within
each $(l_a,l_b)$ group. This can be understood using a semiclassical
picture. In the classical limit, the plane of motion of a particle is
determined by the orbital angular momentum vector ${\bm \ell}$.
Therefore, two particles moving in time-reversed orbits with total
angular momentum zero will have opposite angular momentum vectors. The
larger the value of $l_a, l_b$, the closer the situation will be to
the classical picture and the more aligned the orbital angular
momentum vectors will be (see also
Ref.~\cite[Sect.~2.2]{BrinkBroglia}). Hence, the overlap between the
orbitals is larger with increasing orbital angular momentum. The same
holds with increasing total angular momentum and the larger overlap
leads to the more attractive $\partialwave{1}{S}{0}$ contribution for
larger $j_a$ or $j_b$ within each $(l_a,l_b)$ group.

\subsection{Spin-triplet contributions}
\label{spin1}

The $\partialwave{3}{P}{1}$ partial wave is repulsive at the relevant
energies~\cite{PWA}. However, the contributions to the pairing matrix
elements in Figs.~\ref{sdT1} and~\ref{pfT1} are clearly more
complicated. While the $\partialwave{3}{P}{1}$ contribution is
repulsive between the same spin-orbit partners $(j_>)^2-(j_>)^2$ and
$(j_<)^2-(j_<)^2$ (as expected from phase shifts), we find an
attractive contribution to the pairing matrix elements connecting
$(j_>)^2-(j_<)^2$ and $(j_<)^2-(j_>)^2$ configurations,
where the spin-orbit partners are denoted by $j_>=l+1/2$ and
$j_<=l-1/2$. This effect can be seen prominently in the $pf$ shell in
Fig.~\ref{pfT1}. This is a general property of spin-triplet nuclear
forces and applies also to $T=0, J=1$ pairing matrix elements.

\begin{figure}[t]
\begin{center}
\includegraphics[clip=,scale=0.45]{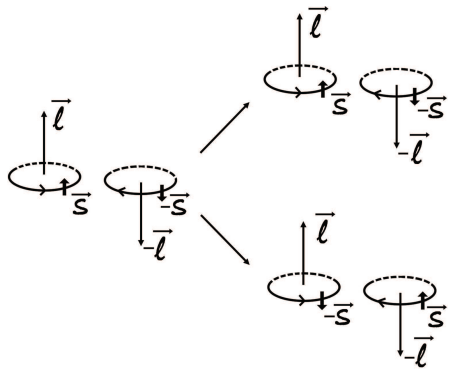}
\end{center}
\caption{Illustration of two particles in time-reversed
states with $j_>=l+1/2$ (left) scattering to the same $(j_>)^2$
configuration (top right) or to the $(j_<)^2$ configuration involving
the spin-orbit partner $j_<=l-1/2$ (bottom right).\label{scatter}}
\end{figure}

The different configurations for scattering of two particles in
time-reversed states are illustrated in Fig.~\ref{scatter}. In the
semiclassical picture, the spin ${\bm s}$ of the particle will be in
the direction or opposite to the orbital angular momentum vector ${\bm
\ell}$ for the $j_>$ or $j_<$ configuration, respectively. If the
two particles scatter from the $j_>$ orbital to its spin-orbit partner
$j_<$ (left to bottom right in Fig.~\ref{scatter}), they will
experience a spin flip that would not occur in the case that the
particles remain in the same $j_>$ orbital (left to top right in
Fig.~\ref{scatter}). This spin flip leads to a relative sign in the
spin part of the wave function $\mid \uparrow \hspace*{-1mm}
\mathcal{T}(\uparrow) \, \ra = \mid \uparrow \downarrow \ra =
\bigl(|S=1\ra + |S=0\ra \bigr)/ \sqrt{2}$ (for the left configuration
in Fig.~\ref{scatter}) versus $\mid \downarrow \hspace*{-1mm}
\mathcal{T}(\downarrow) \, \ra = - \mid \downarrow \uparrow \ra =
\bigl(- |S=1\ra + |S=0\ra \bigr)/ \sqrt{2}$ (for the bottom right
configuration), which results in the relative sign difference of
spin-triplet partial wave contributions to $(j_>)^2-(j_<)^2$ and
$(j_<)^2-(j_>)^2$ versus $(j_>)^2-(j_>)^2$ and $(j_<)^2-(j_<)^2$
configurations. This also holds for $(l_{a\,j_a})^2-(l_{b\,j_b})^2$
matrix elements with $l_a \neq l_b$, leading to an additional
sign for general upper-lower spin-orbit configurations $(j_{a
>})^2-(j_{b <})^2$ and $(j_{a <})^2-(j_{b >})^2$. 

\section{$T=1$ pairing gaps}
\label{T1}

\begin{figure}[t]
\begin{center}
\includegraphics[clip=,scale=0.35]{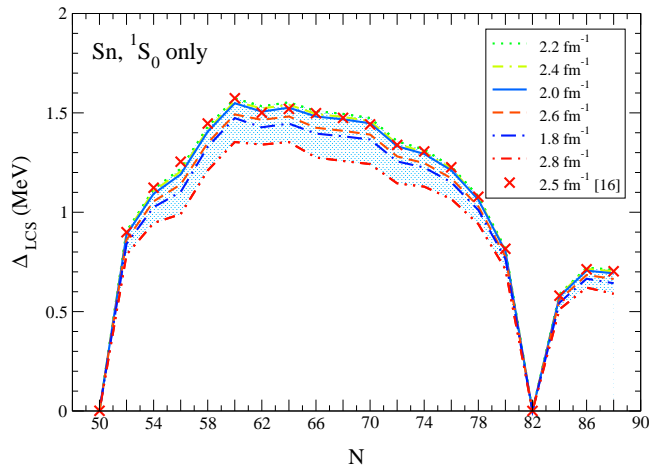}
\end{center}
\caption{Cutoff dependence of the LCS gaps in the tin isotopes
including only the $\partialwave{1}{S}{0}$ contribution to the pairing
interaction. The order in the legend corresponds to the order of the
curves. The crosses denote the results of Ref.~\cite{Lesinski},
based on a smooth $\vlowk$ evolved from the Argonne $v_{18}$ potential
to $\lm=2.5 \fmi$ with $n_{\rm exp}=6$.\label{cutoffdep}}
\end{figure}

As measures of $T=1$ pairing in nuclei, we calculate both
the lowest-quasiparticle-energy canonical state (LCS) gap and
an average gap. The former is defined as the diagonal
pairing gap $\Delta_{aa}$ of Eq.~(\ref{gapeq}), where the LCS
state $a$ is the canonical state with the lowest quasiparticle
energy $E_a$ given by
\be{canonicalqpe}
E_a= \sqrt{(\varepsilon_a - \mu)^2 + \Delta_{aa}^2} \,.
\ee
The canonical basis is defined by diagonalizing the density
matrix, and $\varepsilon_a$ is the canonical single-particle
energy. The LCS state is close to the Fermi level using the 
definition of Eq.~(\ref{canonicalqpe}). We introduce the average
gap as
\be{uvgap}
\overline{\Delta} = \frac{\sum_a \Bigl(\sum_q U^q_aV^q_a\Bigr)
\, (2j_a+1) \: \Delta_{aa}}{\sum_a \Bigl(\sum_q U^q_aV^q_a\Bigr)
\, (2j_a+1)} \,.
\ee
The factor $(\sum_q U^q_aV^q_a) \, (2j_a+1)$ weighs states according
to the degeneracy of the level and averages over an energy window
around the Fermi level, which is approximately given by one shell
above and below the Fermi level (for example, $\approx \mu \pm 10
\mev$ in $\elemA{120}{Sn}$).

We compare the theoretical LCS and average gaps calculated for
even-even nuclei to experimental gaps based on the three-point
mass formula centered on odd nuclei (odd $N$)~\cite{Satula,Bonche},
\be{exp}
\Delta_o^{3}(N) = \frac{(-1)^N}{2} \Big[ B(N+1,Z) - 2 B(N,Z)
+ B(N-1,Z) \Big] \,,
\ee
using experimental binding energies $B(N,Z)$ of Ref.~\cite{masstable}.

For low-momentum interactions, superfluid properties are dominated by
states around the Fermi level and the contributions from the
right-hand side of the gap equation, Eq.~(\ref{gapeq}), fall off as
particle pairs scatter to higher-lying
states~\cite{Hebeler}. Therefore, our results are independent of the
HO basis parameters, as long the HF states around the Fermi level are
reproduced. We have found that $10$ oscillator shells are sufficient
for $6 < \hbar \omega < 18 \mev$. The dependence of the pairing gaps
on the HO basis parameters is at the keV level over this range.

\begin{figure}[t]
\begin{center}
\includegraphics[clip=,scale=0.35]{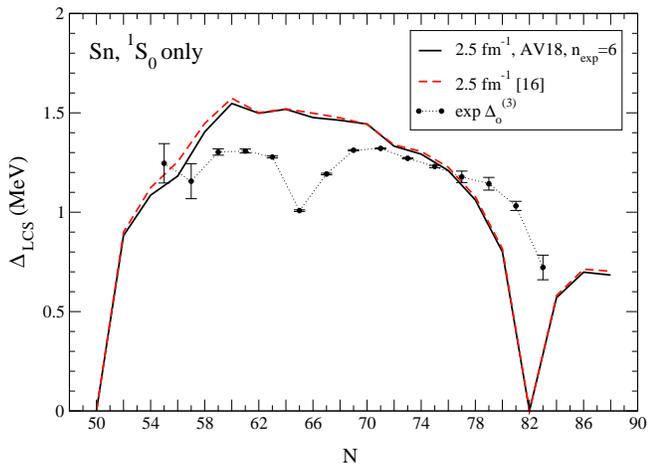}
\end{center}
\caption{Comparison of the LCS gaps in the tin isotopes obtained
using a smooth $\vlowk$ evolved from the Argonne $v_{18}$ potential
to $\lm=2.5 \fmi$ with $n_{\rm exp}=6$ and those of
Ref.~\cite{Lesinski} using the same pairing interaction.
Experimental gaps based on the three-point mass formula are shown for
comparison.\label{check}}
\end{figure}

We first study the cutoff dependence of the LCS gaps in
Fig.~\ref{cutoffdep}. For simplicity, we here include only the
$\partialwave{1}{S}{0}$ contribution. We find that the cutoff
variation is small overall and for $2.0 \fmi \lesssim \lm \lesssim 2.5
\fmi$ the gaps are approximately cutoff independent.  Next, we check
our results against the gaps obtained in Ref.~\cite{Lesinski} using
the same pairing interaction: the $\partialwave{1}{S}{0}$ part of the
smooth $\vlowk$ evolved from the Argonne $v_{18}$ potential to
$\lm=2.5 \fmi$ with $n_{\rm exp}=6$ (and the same Skyrme functional
SLy4 in the particle-hole channel). The very good agreement is shown
in Fig.~\ref{check}. The gaps are within $0.5 \%$ for almost all tin
isotopes, with a couple of them ($N=54$ and $56$) showing a larger (but
still small) difference of $3 \%$. This demonstrates that the
truncation after the first HFB iteration introduces a negligible error
for pairing gaps. In the following, all results are based on the
RG-evolved N$^3$LO potential of Ref.~\cite{N3LO} with $\lm = 2.0 \fmi$
and $n_{\rm exp}=4$.

\begin{figure}[t]
\begin{center}
\includegraphics[clip=,scale=0.35]{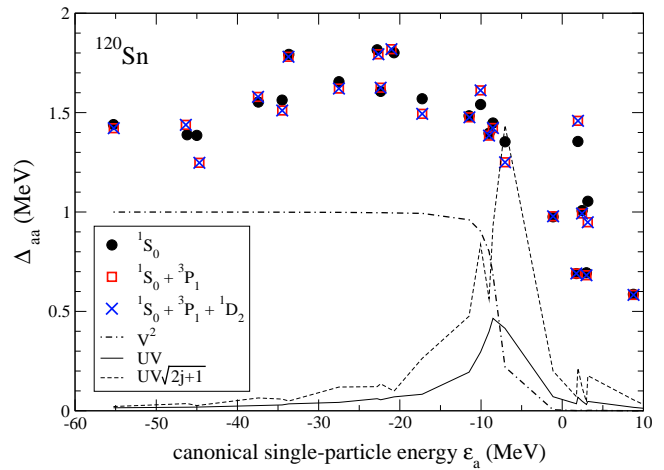}
\end{center}
\caption{State-dependent canonical pairing gap $\Delta_{aa}$ in
$\elemA{120}{Sn}$ versus canonical single-particle
energy $\varepsilon_a$. In addition, we show the occupation factor
$\sum_q V^q_aV^q_a$ (labeled $V^2$), the $\sum_q U^q_aV^q_a$ factor
(labeled $UV$), and the $(\sum_q U^q_aV^q_a) \sqrt{2j_a+1}$ factor
(labeled $UV \sqrt{2j+1}$) that enters the gap equation, 
Eq.~(\ref{gapeq}). The $y$-axis for the different factors
is dimensionless and the
factors corresponding to the three partial-wave cases are
the same on this scale.\label{statedep}}
\end{figure}

\begin{figure}[t]
\begin{center}
\includegraphics[clip=,scale=0.345]{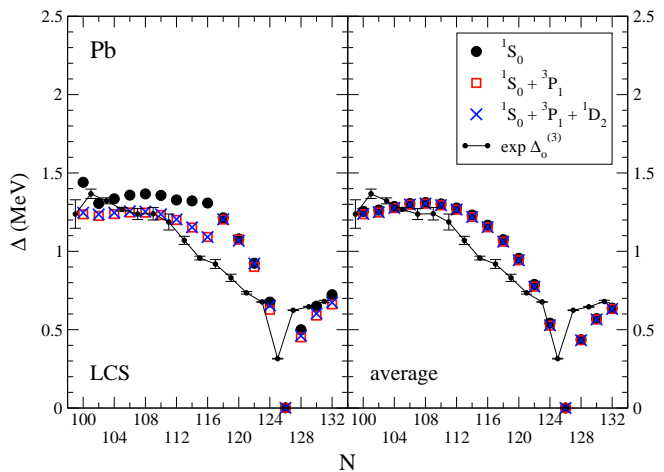}
\end{center}
\caption{LCS and average gaps in the lead isotopes with increasing
partial-wave contributions to the pairing interaction. Experimental
gaps based on the three-point mass formula are shown for comparison.
\label{Pb}}
\end{figure}

In Fig.~\ref{statedep}, we plot the state dependence of the canonical
pairing gap $\Delta_{aa}$ versus the canonical single-particle energy
$\varepsilon_a$. The occupation factor ($V^2$) and the $UV$ factors
are included to show the Fermi level. As expected from the pairing
matrix elements, we observe that the impact of the
$\partialwave{3}{P}{1}$ partial wave is attractive for some states
(pairing gap is increased) and repulsive for others (pairing gap is
decreased). This is due to the interplay between upper-lower
spin-orbit configurations for spin-triplet contributions to pairing,
discussed in Sect.~\ref{spin1}. Given the $(\sum_q U^q_aV^q_a)
\sqrt{2j_a+1}$ factor shown in Fig.~\ref{statedep}, the effect of the
$\partialwave{3}{P}{1}$ partial wave on $\Delta_{aa}$ in the gap
equation, Eq.~(\ref{gapeq}), is driven by the sign of the
$\partialwave{3}{P}{1}$ contribution to the pairing matrix elements $
\la a \, a \, | \, (1 - P_{12}) \, \vlowk \, | \, f \, f \, \ra_{J=0,
T=1, T_z=-1}$, where $f$ denotes the LCS state close to the Fermi
level where $(\sum_q U^q_aV^q_a) \sqrt{2j_a+1}$ is peaked. Following
the analysis of Sect.~\ref{spin1}, for the LCS state ($a=f$) or if $a$
and $f$ are the same upper or lower spin-orbit configurations, the
$\partialwave{3}{P}{1}$ partial wave will decrease the state-dependent
pairing gap $\Delta_{aa}$, while for different spin-orbit
states $a$, the $\partialwave{3}{P}{1}$ contribution will increase
$\Delta_{aa}$.  Moreover, as expected from Figs.~\ref{sdT1}
and~\ref{pfT1}, the impact of higher partial waves beyond the
$\partialwave{3}{P}{1}$ channel is small in Fig.~\ref{statedep} and
for $T=1$ pairing gaps in other nuclei.

\begin{figure}[t]
\begin{center}
\includegraphics[clip=,scale=0.345]{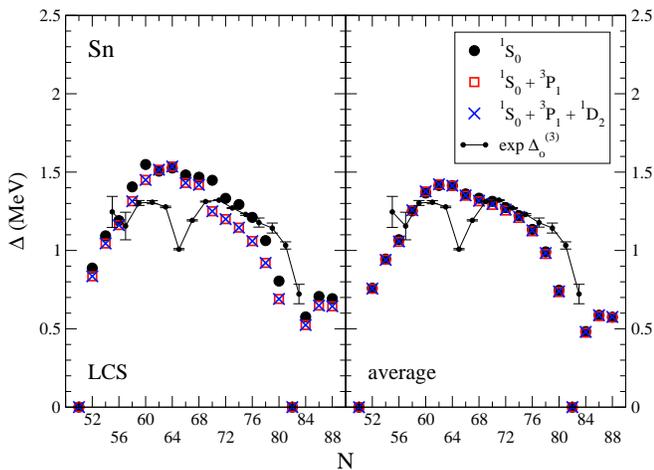}
\end{center}
\caption{Same as Fig.~\ref{Pb}, but for tin isotopes.\label{Sn}}
\end{figure}

\begin{figure}[t]
\begin{center}
\includegraphics[clip=,scale=0.345]{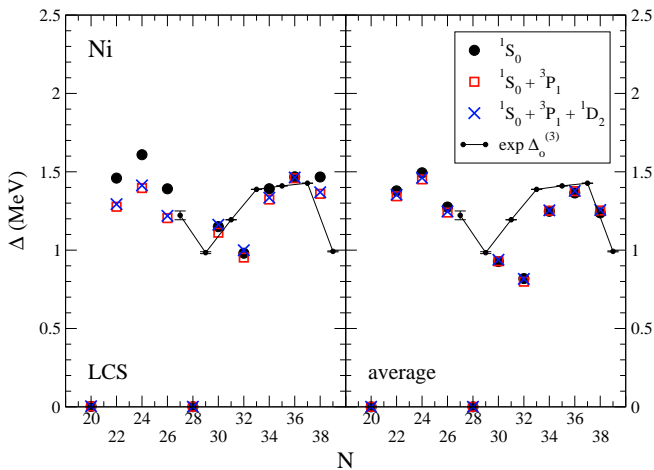}
\end{center}
\caption{Same as Fig.~\ref{Pb}, but for nickel isotopes.\label{Ni}}
\end{figure}

\begin{figure}[t]
\begin{center}
\includegraphics[clip=,scale=0.345]{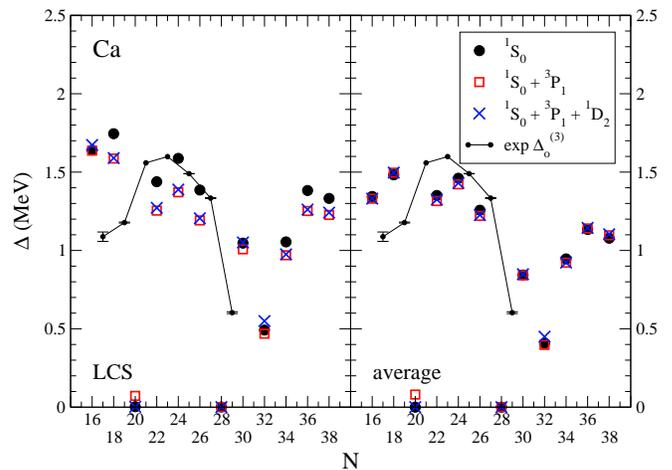}
\end{center}
\caption{Same as Fig.~\ref{Pb}, but for calcium isotopes.\label{Ca}}
\end{figure}

\begin{figure}[t]
\begin{center}
\includegraphics[clip=,scale=0.345]{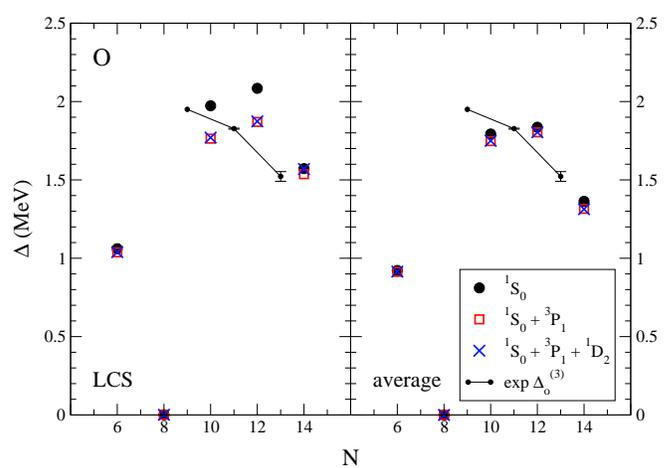}
\end{center}
\caption{Same as Fig.~\ref{Pb}, but for oxygen isotopes.\label{O}}
\end{figure}

We present a more global study of neutron-neutron pairing gaps by
calculating the LCS and average gaps for even-even nuclei in the lead,
tin, nickel, calcium and oxygen isotopes in Figs.~\ref{Pb}-\ref{O}.
The theoretical gaps are shown with increasing partial-wave
contributions to the pairing interaction and in comparison to the
experimental gaps based on the three-point mass formula,
Eq.~(\ref{exp}), centered on the neighboring odd nuclei. Higher
partial waves were also included in Ref.~\cite{Hergert}. The
additional contributions beyond the standard $\partialwave{1}{S}{0}$
channel are dominated by the $\partialwave{3}{P}{1}$ partial
wave. They lead to a decrease of the LCS gaps of up to $15 \%$ and can
change the isotopic dependence at this level. On the other hand,
because of the interplay of spin-orbit configurations for the
spin-triplet $\partialwave{3}{P}{1}$ channel and the resulting
alternation of repulsive and attractive contributions to the
state-dependent pairing gap in Fig.~\ref{statedep}, we find that the
average gap changes little when additional partial waves beyond the
$\partialwave{1}{S}{0}$ channel are included. This shows that the
impact of the $\partialwave{3}{P}{1}$ contribution depends on the
definition of theoretical gap. We therefore conclude that future
studies should compare directly calculated odd-even mass differences
to the experimental gaps. This has recently been carried out in
Ref.~\cite{Duguet2} using only the $\partialwave{1}{S}{0}$
contribution to the pairing interaction.

\section{$T=0, J=1$ pairing matrix elements}
\label{T0}

\begin{figure}[t]
\begin{center}
\includegraphics[clip=,scale=0.35]{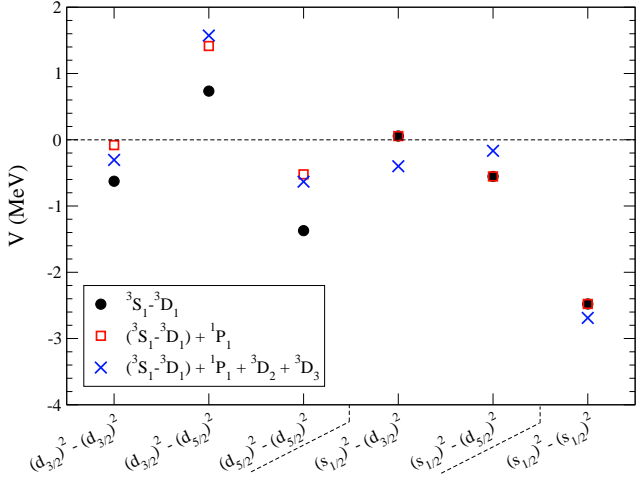}
\end{center}
\caption{$T=0, J=1$ pairing matrix elements in the $sd$ shell based on
the same RG-evolved N$^3$LO potential as used to calculate the $T=1,
J=0$ pairing matrix elements in Fig.~\ref{sdT1}.\label{sdT0}}
\end{figure}

\begin{figure}[t]
\begin{center}
\includegraphics[clip=,scale=0.35]{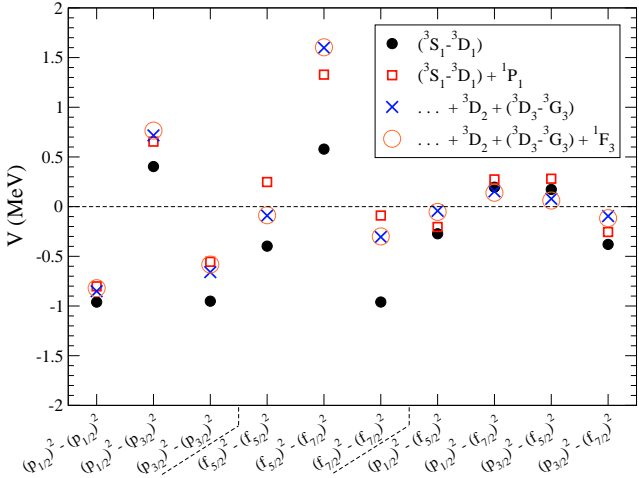}
\end{center}
\caption{Same as Fig.~\ref{sdT0} but for the $pf$ shell, where partial
waves up to the $\partialwave{3}{G}{3}$ channel can contribute to
$T=0, J=1$ pairing matrix elements.\label{pfT0}}
\end{figure}

We turn to the case of neutron-proton $T=0, J=1$ pairing, with a focus
on the relative pairing strengths compared to the $T=1, J=0$
channel. In Figs.~\ref{sdT0} and~\ref{pfT0}, we show the $jj$-coupled
$T=0, J=1$ pairing matrix elements in the $sd$ and $pf$ shell. The
matrix elements are in the HO basis with $\hbar \omega = 10 \mev$, but
the general effects of partial waves are qualitatively similar for
other values of $\hbar \omega$ or in the HF basis. While all $T=1,
J=0$ pairing matrix elements are attractive at the S-wave level, for
$T=0$ pairing the S-wave contribution is given by the spin-triplet
$\partialwave{3}{S}{1} - \partialwave{3}{D}{1}$ channel, which is
influenced by the interplay between spin-orbit configurations discussed in
Sect.~\ref{spin1}. As a result, although the spin-triplet S-wave
interaction is attractive, pairing matrix elements connecting
different spin-orbit configurations, $(j_{a >})^2-(j_{b <})^2$ and
$(j_{a <})^2-(j_{b >})^2$, are repulsive. This can be seen clearly in the
$(d,d)$, $(p,p)$, $(f,f)$ and $(p,f)$ groups at the $\partialwave{3}{S}{1}
- \partialwave{3}{D}{1}$ level in Figs.~\ref{sdT0} and~\ref{pfT0}.

The next partial-wave contribution is due to the repulsive
$\partialwave{1}{P}{1}$ spin-singlet channel (in contrast to the
spin-triplet $\partialwave{3}{P}{1}$ for $T=1$ pairing), so that it
decreases the attractive pairing strength of all $T=0, J=1$ pairing
matrix elements. Finally, Figs.~\ref{sdT0} and~\ref{pfT0} show smaller
contributions due to the spin-triplet D waves, again with alternating
attractive and repulsive character.

\begin{figure}[t]
\begin{center}
\includegraphics[clip=,scale=0.35]{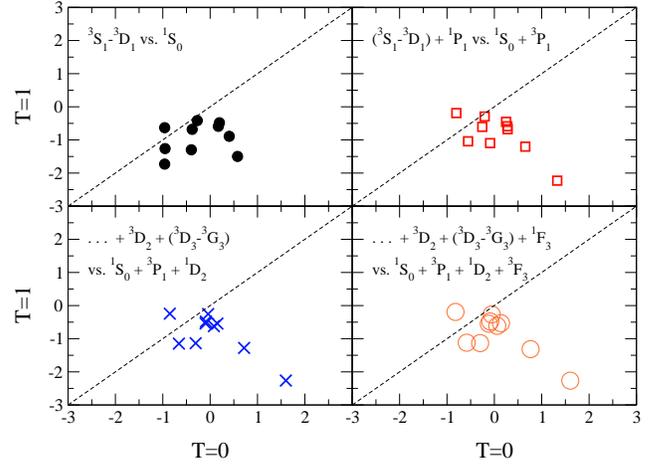}
\end{center}
\caption{Comparison of $T=1, J=0$ versus $T=0, J=1$ pairing matrix
elements in the $pf$ shell. Interactions that favor $T=1$ pairing
lie below the diagonal. The only point above the diagonal
corresponds to the $(p_{1/2})^2-(p_{1/2})^2$ matrix element. The
pairing matrix elements and the symbols used are the same as in
Figs.~\ref{pfT1} and~\ref{pfT0}.\label{T0vsT1}}
\end{figure}

We compare the pairing strengths in the $pf$ shell by plotting the
$T=1, J=0$ versus $T=0, J=1$ pairing matrix elements in
Fig.~\ref{T0vsT1}. This clearly demonstrates that nuclear forces in
the $pf$ shell favor $T=1$ over $T=0$ pairing, except for the
$(p_{1/2})^2-(p_{1/2})^2$ matrix elements. This is in contrast with
the expectation based on the relative S-wave interactions in free
space, which favor the $T=0$ over $T=1$ channel, with a bound deuteron
compared to the nearly-bound state in the $\partialwave{1}{S}{0}$
channel.

\subsection{Spin-orbit suppression}
\label{LS}

\begin{figure}[t]
\begin{center}
\includegraphics[clip=,scale=0.35]{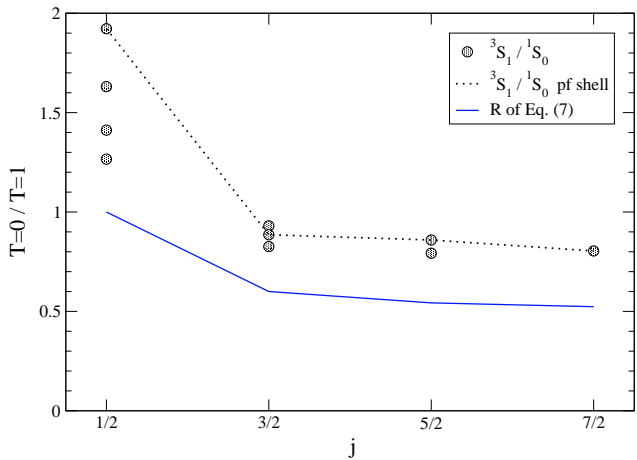}
\end{center}
\caption{Ratio of the diagonal $(l_{a\,j_a})^2-(l_{a\,j_a})^2$
pairing matrix elements in the $T=0, J=1$ over the $T=1, J=0$
channel. Results are shown including only S waves as a function of
the total angular momentum $j$ up to the $pf$ shell. The
dotted line connects the $pf$-shell matrix elements. The solid line
is the ratio $R$ of Eq.~(\ref{ratio}) for S-wave contact
interactions with identical strengths.\label{T0vsT1ratio}}
\end{figure}

In addition to the effects of the repulsive $\partialwave{1}{P}{1}$
channel and of D waves, the suppression of $T=0$ pairing at the S-wave
level in Fig.~\ref{T0vsT1} is due to spin-orbit splitting between
single-particle states. The impact of spin-orbit splitting on $T=0$
pairing properties has also been pointed out in shell-model
calculations using empirical interactions~\cite{SM1,SM2}. In our
analysis, the significance of spin-orbit splitting lies in the choice of
the $jj$-coupled basis and in our assumption that pairing matrix
elements based on low-momentum interactions provide a first
approximation for pairing in a $j$-shell~\cite{Duguet1,Lesinski}.

We discuss this first at the level of the $\partialwave{1}{S}{0}$ and
$\partialwave{3}{S}{1}$ partial waves. The corresponding phase
shifts~\cite{PWA} (and also low-momentum
interactions~\cite{Vlowk,smooth}) have very similar momentum
dependences, because at low energies nuclear forces mainly differ by
providing slightly more attraction in the $T=0$ channel to lead to a
loosely bound deuteron. The similarity of the $\partialwave{1}{S}{0}$
and $\partialwave{3}{S}{1}$ channels is also reflected in large
scattering lengths, $a_{^1\text{S}_0} = -23.768 \, {\rm fm}$ and
$a_{^3\text{S}_1} = 5.420 \, {\rm fm}$. At low energies, nuclear
interactions are therefore close to Wigner's SU(4) limit, where S-wave
contact interactions have identical strengths~\cite{Mehen}.

In order to study the ratio $R$ of the $T=0, J=1$ over the $T=1, J=0$
pairing strengths, we consider two S-wave contact interactions
with the same $T=1$ and $T=0$ strengths. The resulting pairing
matrix elements can be calculated analytically (see, for example,
Ref.~\cite{Heyde}) and the ratio $R$ is given by a geometrical factor:
\begin{align}
R &= \frac{\la n_a lj \, n_a lj \, | \, (1 - P_{12}) \, V_{^3\text{S}_1} \, 
| \, n_b lj \, n_b lj \, \ra_{J=1, T=0}}{\la n_a lj \, n_a lj \, | \,
(1 - P_{12}) \, V_{^1\text{S}_0} \, | \, n_b lj \, n_b lj \,
\ra_{J=0,T=1}} \,, \nonumber \\[1mm]
&= \frac{\left( \begin{array}{ccc}
j & j & 1 \\
1/2 & -1/2 & 0 \end{array} \right)^2
+ \left( \begin{array}{ccc}
j & j & 1 \\
1/2 & 1/2 & -1 \end{array} \right)^2}{
\left( \begin{array}{ccc}
j & j & 0 \\
1/2 & -1/2 & 0 \end{array} \right)^2} \,, \nonumber \\[1mm]
&= \frac{4j^2 + 4j +3}{8j(j+1)} \,,
\label{ratio}
\end{align}
where $(\ldots)$ are 3j symbols~\cite{angular}.
We find $R = 1$ for $j=1/2$, $R = 1/2$ in the limit of large $j$, and
already $R \leqslant 3/5$ for $j \geqslant 3/2$~\cite{T0puzzle}. 
This geometrical factor is due to the spin-orbit suppression and provides 
a simple explanation why S-wave nuclear forces favor $T=1, J=0$ over $T=0, J=1$
pairs, except in $j=1/2$ orbitals.

\begin{figure}[t]
\begin{center}
\includegraphics[clip=,scale=0.35]{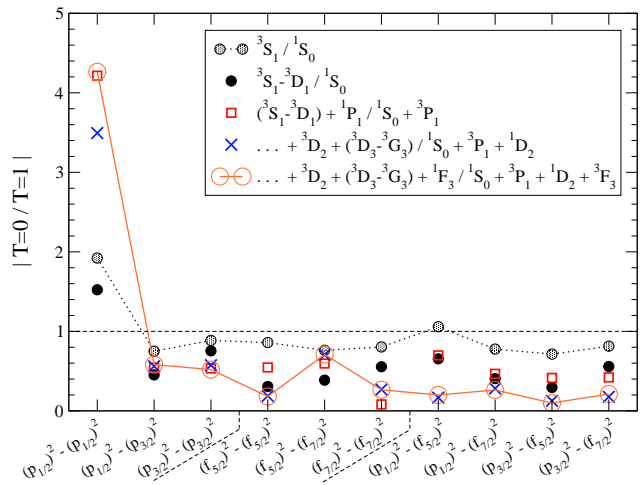}
\end{center}
\caption{Absolute value of the ratio of $T=0, J=1$ over the $T=1, J=0$
pairing matrix elements (all $(l_{a\,j_a})^2-(l_{b\,j_b})^2$) in the
$pf$ shell. The anomalously large ratio for the
$(p_{1/2})^2-(p_{1/2})^2$ matrix element beyond S waves is due to the
small $T=1, J=0$ pairing matrix element in this case (see
Fig.~\ref{pfT1}).\label{absratio}}
\end{figure}

In Fig.~\ref{T0vsT1ratio}, we compare the ratio $R$ to the diagonal
$(l_{a\,j_a})^2-(l_{a\,j_a})^2$ pairing matrix elements up to the $pf$
shell. We find that the trend of Eq.~(\ref{ratio}) agrees nicely. As
expected, the ratio is somewhat larger for low-momentum interactions,
because the $\partialwave{3}{S}{1}$ part is more attractive than the
$\partialwave{1}{S}{0}$ partial wave. For all $j \geqslant 3/2$, the
$T=1, J=0$ channel is favored. Finally, we include all partial-wave
contributions and show in Fig.~\ref{absratio} the absolute value of
the ratio of $T=0, J=1$ over the $T=1, J=0$ pairing matrix elements in
the $pf$ shell. This demonstrates that nuclear forces favor $T=1, J=0$
over $T=0, J=1$ pairing, except in low-$j$ orbitals.

The suppression $R=1/2$ in the limit of large $j$ can be interpreted
as a distribution of the $\partialwave{3}{S}{1}$ strength on more
configurations, compared to the $\partialwave{1}{S}{0}$ case, because
there are only two states for $J=0$: $(j_>)^2$ and $(j_<)^2$; compared
to four states for $J=1$: $(j_>)^2$, $(j_<)^2$, $(j_> j_<)$ and $(j_<
j_>)$. In addition, a semiclassical picture for the suppression of
$T=0$ pairing has been developed in Ref.~\cite{Simone}.

\section{Conclusions and outlook}
\label{concl}

We have studied in detail partial-wave contributions of nuclear forces
to pairing in nuclei, at the level of the pairing matrix elements and
for neutron-neutron pairing gaps in semi-magic isotopic chains.  For
$T=1, J=0$ pairing, the repulsive $\partialwave{3}{P}{1}$ channel
decreases the LCS gaps by up to $15 \%$ compared to the gaps obtained
from the standard $\partialwave{1}{S}{0}$ contribution, while the
changes in the average gap were found to be small. The latter is due
to the alternation of repulsive and attractive contributions for
spin-triplet nuclear forces between paired nucleons. While we have
focused on neutron-neutron pairing gaps, our conclusions equally apply
to proton-proton gaps.

We expect the $\partialwave{3}{P}{1}$ effects will be more important
for phenomena involving state-dependent pairing gaps, such as
particle-transfer reactions. From the differences between the LCS and
average gaps beyond the $\partialwave{1}{S}{0}$ channel, we conclude
that future studies should compare directly calculated odd-even mass
differences to the experimental gaps (see also
Ref.~\cite{Duguet2}). Finally, important future work for pairing gaps
is to include the effects of three-nucleon forces (for the impact in
neutron matter, see Ref.~\cite{neutmatt}), many-body contributions or
induced interactions~\cite{Barranco,Pastore}, and center-of-mass
corrections to pairing interactions~\cite{Hergert}.

Based on the comparison of pairing matrix elements, we have shown that
nuclear forces favor $T=1, J=0$ over $T=0, J=1$ pairing, except in
low-$j$ orbitals. This is in contrast to the relative S-wave strengths
in free space. We have traced the suppression of $T=0$ pairing to two
origins. First is the spin-orbit splitting, which results in the
$\partialwave{3}{S}{1}$ strength being distributed on more
configurations, compared to the $\partialwave{1}{S}{0}$ strength in
the case of $T=1$ pairing. In the SU(4) limit where both relative
S-wave interactions are equally attractive, this leads to a relative
suppression of the pairing interaction for large-$j$ orbitals by a
factor $1/2$, in favor of the $T=1, J=0$ case. The deuteron
$\partialwave{3}{S}{1}$ channel would have to be twice as attractive
as $\partialwave{1}{S}{0}$ to overcome this spin-orbit
suppression. Second, $T=0$ pairing is weakened by the additional
repulsive $\partialwave{1}{P}{1}$ channel and by the effects of D
waves, more strongly compared to the effects of higher partial waves
in the $T=1$ pairing case. Our findings render the standard free-space
motivation for $T=0$ pairing at best incomplete.

\acknowledgments

We are grateful to G.\ F.\ Bertsch for discussions on $T=0$ pairing 
and also thank T.\ Duguet, K.\ Hebeler, V.\ Koch, T.\ Lesinski and A.\ Pastore 
for useful conversations.
This work was supported in part by the UNEDF
SciDAC Collaboration under DOE Grant DE-FC02-07ER41457, the NSF under
Grant 0835543, the DOE under Contract No. DE-AC02-05CH11231 (LBNL),
the Natural Sciences and Engineering Research Council of Canada
(NSERC) and the Helmholtz Alliance Program of the Helmholtz
Association, contract HA216/EMMI ``Extremes of Density and
Temperature: Cosmic Matter in the Laboratory''. TRIUMF receives
funding via a contribution through the National Research Council
Canada.

\bibliographystyle{apsrev}

\end{document}